# Mechanism and kinetics of phase transitions and other reactions in solids

Yuri Mnyukh
*76 Peggy Lane, Farmington, CT, USA, e-mail: yuri@mnyukh.com*
(Dated: October 7, 2011)

The work is presented, leading to the universal *contact* molecular mechanism of phase transitions and other reactions in solid state. The two components of the mechanism - nucleation and interface propagation - are investigated in detail and their role in the kinetics is specified. They were shown to be peculiar: nucleation is "pre-coded", rather than resulted from a successful fluctuation, and the interface propagates by molecular filling of thin layers in the transverse direction. The structure of the nucleation sites is determined. The inherent instability and irreproducibility of the kinetics in question is revealed. A *linear kinetics*, as opposed to the *bulk kinetics*, is introduced and shown to be in accord with the *contact* mechanism. Ferromagnetic phase transition and magnetization are added to the list of solid-state reactions; neither occurs without structural rearrangement.

## 1. Introduction

The recent set of articles summarizing basics of solid-state phase transitions [1-7] will be supplemented here by analysis of their kinetics − an aspect of essential theoretical and applied importance. The term *kinetics* means relationships between the macroscopic rate of a phase transition and any conditions or parameters it depends on. The notion *kinetics* implies *phase coexistence*, for it makes sense only if the mass fraction between the two phases is changing. The only conceivable way of this change is *nucleation and propagation of interfaces*. In other words, investigation of kinetics of phase transitions already means recognition of their nucleation-and-growth mechanism. Thermodynamics allows only two mechanisms of crystal phase transitions: (a) by nucleation-and-growth and (b) by instant change at a critical point [1]. If a phase transition is a "critical phenomenon", its kinetics must not exist, for it comprises all the matter at once as soon as its critical temperature $T_c$ is attained. The rate of this instant transition is function neither time *t*, nor temperature T. The concepts "kinetics" and "critical phenomenon" are incompatible. It should be noted that instant phase transitions have not been found [1,3].

Kinetics of phase transitions is inseparable from their molecular mechanism. As soon as the nucleation-and-growth nature of solid-state phase transitions is recognized, it becomes evident that comprehension of their kinetics requires a certain knowledge of the structure and properties of the nuclei and interfaces. Critical step in that direction has been a discovery of *edgewise* (or *stepwise*) molecular mechanism of phase transitions [8-10]. The nucleation, in particular, was found quite different from the "classical" interpretation, its features critically affecting the phase transition kinetics (see below).

Closely preceding to that discovery was the 1960 *International Symposium on Reactivity of Solids* [11]. It "focused [its] attention on the mechanism and kinetics of reaction in solids". It dealt with such reactions as polymorphic phase transitions, recrystallization, decomposition, chemical reduction and polymerization, not counting those involving also liquid or gas. The vast literature that treated almost all phase transitions as "continuous" and "critical phenomenon" was ignored as if not existent. It was observed that "imperfections are preferred sites for internal nucleus formations". (The term "preferred", however, does not eliminate a possibility of homogeneous nucleation). These imperfections are: vacancies, interstitials, foreign atoms or ions, linear dislocations, screw dislocations. Their interaction and diffusion were discussed. It was noted that the more perfect crystal, the smaller is its reactivity. The term "nucleation and growth" was common. It was not new, however, for in 1930's and 1940's it was used in developing of what can be called "bulk kinetics" theories. Nevertheless, nucleation-and-growth phase transitions were not considered the only way to occur. For example, one contributor [12] classified solid-state transformations as *topotactic*, *epitactic* and *reconstructive*, only the last one to occur by nucleation and growth. In another case [13] the imaginary way of the olivine-spinel restructuring through intermediate states, assisted by dislocations and diffusion, was proposed. It was not mentioned how the notion "kinetics" can be applied to the phase transitions that somehow occur without nucleation and growth.

In the *bulk kinetics* the mass fraction *m* of one phase in a two-phase specimen was the value of interest. Its rate depends on both nucleation and growth in unknown proportion, different in every particular case. Nucleation critically depends on the presence, distribution and generation of specific lattice defects. It is quite different in a perfect and imperfect single crystal, in a big and small crystal, in a single crystal and polycrystal, in a fine-grain and coarse-grain polycrystal or powder. Growth is not a stable value either, not repeating itself, for example, in cycling phase transitions. The nucleation and growth, when they act together, are not only irreproducible and uncontrollable, there is no way to theoretically separate their contributions in order to calculate the total bulk rate. In this context, the theoretical approach called "formal kinetics" should be mentioned where an attempt to separate them was undertaken. The Avrami work [14] is most known. One of his main assumptions − isothermal rate of nucleation − was invalid due to a "pre-coded" character of both nucleation sites and the nucleation temperatures [15]. Bulk kinetics can shed no light upon the physics of phase transitions or properly account for their kinetics. As a minimum condition, the nucleation and interface motion contributions must be experimentally separated. The best way to do that is a visual observation of nuclei formation and interface motion in optically transparent single crystals. This is a method of *interface kinetics*. The purpose is to reveal its physics, rather then phenomenology.

As mentioned, there is a reproducibility problem. The absolute velocity V of interface motion is not reproduced even in the same single crystal. Fortunately, valuable information can be obtained from *relative* V changes. As will be shown, it helps not only verify and substantiate the *contact* mechanism, but also to penetrate deeper into its details. It will be demonstrated that the universal *contact* mechanism easily accounts for all the complexity, versatility, and poor reproducibility of the kinetics of solid-solid phase transitions..

## 2. Interface and its motion

After the discovery has been made that phase transitions in single crystals of *p*-dichlorobenzene (PDB) and other substances were a growth of well-bounded crystals of the new phase (Fig. 1) [16-19] and that orientation relationship (OR) between the initial and resultant crystals did not exist [20], the attention was concentrated on the mode of interface propagation. Observation of the interface was undertaken under maximum resolution attainable with a regular optical microscope. It was found that its motion has *edgewise* (or "*stepwise*") mechanism [8-10]. Its advancement in the normal direction proceeded by transverse shuttle-like strokes of small steps (kinks), every time adding a thin layer to it (Fig. 2). That was the same mechanism of crystal growth from liquids and gases [21,22], going down to the molecular level. A generalization came to light: any process resulting in a crystal state, whichever the initial phase is − gas, liquid, or solid − is a *crystal growth*. It proceeds by the edgewise molecule-by-molecule formation of layers and layer-by-layer additions to the natural crystal face. While crystal growth from gaseous and liquid phases is called *crystal growth*, and crystal growth from a solid phase is called *phase transformation* or *phase transition*, the difference is semantic. This does not mean, of course, that crystal growth in a crystal medium does not have its specificity.

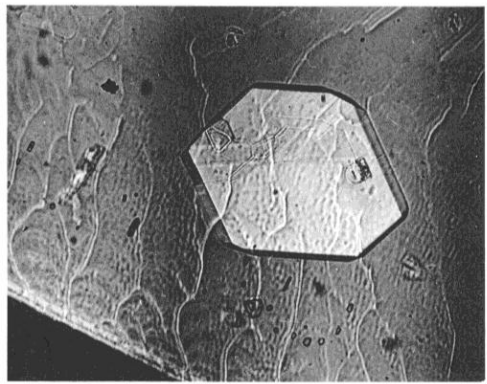

Fig. 1. Growth of well-bounded H single crystal in the single-crystalline PDB plate. A part of the natural edge of the L crystal is visible at the left lower corner. The real diameter of the H crystal is 0.4 mm. (L and H are low- and high-temperature phases respectively).

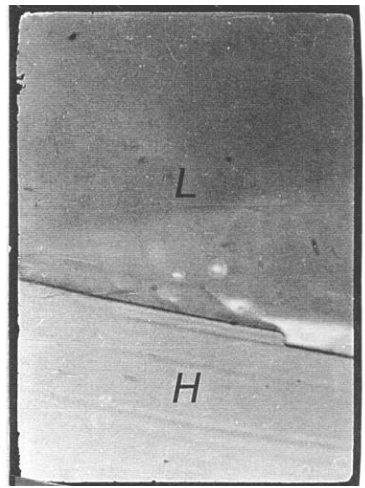

Fig. 2. L → H phase transition in PDB. The 2.5 μm high "kink" (step) is moving from left to right. The kink is, in fact, an avalanche of molecular steps.

The physical model of a solid-solid interface and the manner of molecular rearrangement at this interface was needed. A model assuming existence of a vapor gap between the phases was tried and ended in impasse by Hartshorne and colleagues [23-35]. They concluded that the activation energies of solid-state phase transitions $E_A$ and the heat of sublimation $E_S$ were equal, but the velocity of interface movement was $10^3$ to $10^5$ times higher than can be provided by evaporation into the gap ("Hartshorne's paradox"). On the other hand, any amorphous interlayer of excited molecules cannot exist either: phase transition is localized only at the kinks, and there would be plenty of time for the excited molecules, if any, of the smooth interface to join the stable phase.

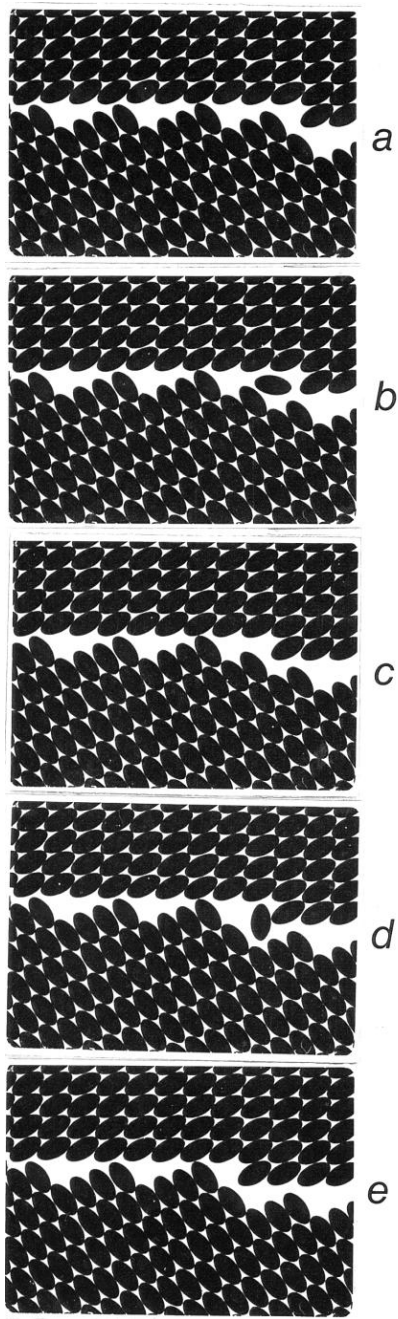

Fig. 3. Molecular rearrangement at the *contact* interface during phase transition (frames of the animated film). The effective gap between the phases is 0.5 molecular layer. Building up a new molecular layer (a → e).

The *contact* interface, shown in Fig.3 with a 2-D model, meets all the observations. There is neither vacuum, nor amorphous transitional layer. Two crystal phases are simply in contact with each other, coupled by the molecular forces. The *contact* structure of the interface requires neither any correlation between the structural parameters of the two phases, nor a particular OR. Interface on the side of the resultant crystal consists of molecules closely packed into a layer of low crystallographic indices $(h, k, \ell)$. In relation to the initial lattice this direction can be irrational. There is a net of microcavities at the interface that cannot accommodate additional molecules. No essential lattice distortions exist.

The *contact* interface was subjected to a number of additional tests [36,9]:

♦ *Coupling at the contact interface (observation).* While the Hartshorne's teem worked with polycrystal films, our dealing with single crystals allowed immediately reject the "vapor gap" model. Indeed, handling a "hybrid" crystal crossed by a flat interface did not result in its separation into two detached parts, as it would be in the case of a vapor gap between the phases. Consequently, there is an essential coupling at the interface. On the other hand, the interface is the weakest section, as expected from the its *contact* model. Application of an external force breaks the "hybrid" crystal exactly along the interface into two separate individual phases.

♦ *Coupling at the contact interface (measurement and calculation).* The coupling force between the phases at their interface was measured by breaking the "hybrid" crystal by a microdynamometer and then compared with the calculated value for the phases separated by 0.5 molecular layer as suggested by the model. The two values turned out to be in a reasonable agreement.

♦ *Solution of the Hartshorne's paradox.* Molecular rearrangement at the *contact* interface is a one-step molecular relocation under the attractive action of resultant phase. This process can be termed "stimulated sublimation". Accordingly, $E_A$ is lower than $E_S$ by the attraction energy $E_{attr}$. For the effective gap 0.5 molecular layer, the computer calculations led to

$E_A = \sim 0.7\ E_S$. When believing that $E_A = E_S$, the Hartshorne's team was not far from truth, but the 30% difference accounts for $\sim 10^4$ times faster phase transition as compared to sublimation [37].

Fig. 3 also shows how the interface advances by molecule-by-molecule process. The frames sequence illustrates the edgewise movement of a molecular step. Molecules detach from one side to build up a closely packed layer on the opposite side. Once one layer is completed, the interface becomes advanced by one interlayer spacing, while the "contact" structure of the interface at its new position is preserved. For the phase transition to continue, a new nucleus has to form on the interface. Here the specificity comes about. Formation of a 2-D nucleus on the flat regions of the contact interface requires extra free space, such as a vacancy or, possibly, their cluster. If it is available, a new molecular step can form to run along the interface. This vacant space will accompany the running step, providing for steric freedom just where the molecular relocation is to occur at the moment. The vacant space will ultimately come out on the crystal surface. Formation of a new step will require another vacant space residing at the new interface position or migrating to it. The *presence of these crystal defects in sufficient quantity is a necessary condition for a phase transition to proceed*. Their availability and diffusion are major controlling factors of the phase transition kinetics.

### 3. Nucleation in crystals

Nucleation is one of the two elements - nucleation and growth - constituting the phenomenon of solid-state phase transitions. Nevertheless, there is vast theoretical literature that treats them as if the phenomenon of nucleation (and subsequent growth) is nonexistent. For example, nucleation is missing in the three (unrelated) books *Structural Phase Transitions* [38-40], in many other books on the subject [*e. g*., 41-48], in most volumes of *Phase Transitions and Critical Phenomena* [49], and in innumerous journal articles. Yet, some literature on nucleation in a solid state exists, mainly owing to needs of solid-state reactivity, as described above, and physical metallurgy [50.51]. In the latter case the theory of nucleation in a liquid phase was slightly modified to cover solid state [50,52]. But no one previously experimentally verified its theoretical assumptions by using the simplest and most informative objects − good quality small transparent single crystals.

The following data were accumulated with temperature-induced phase transitions in tiny (1 - 2 mm) good quality transparent organic single crystals under controlled temperatures, PDB being the main object [15]. Notations: L is low-temperature phase, H is high-temperature phase; $T_o$ is temperature when free energies of the phases are equal; $T_m$ is temperature of melting; $T_n$ is temperature of nucleus formation. In PDB, $T_o = 30.8$ °C and $T_m = 53.2$ °C.

- *Nucleation requires overheating / overcooling*. Nucleation never occurs at $T_o$ or in a certain finite vicinity of it. The "prohibited" range for PDB is at least 28 to 32 °C. Upon slow heating, nucleation in most PDB crystals will not occur until the temperature exceeds 38 °C. Often L phase melts at $T_m$ without transition into H.

- *Nucleation is a rare event*. Slow heating (*e. g*., 1 to 10 °C per hour) does not produce many nuclei. Usually there are only a few units, or only one, or no nucleation sites at all. This observation is at variance with the notion "rate of nucleation" and the statistical approach to nucleation - at least as applied to a single-crystal medium.

- *Exact temperature of nucleation $T_n$ is unknown a priori*. Formation of a nucleus upon slow heating of a PDB single crystal will occur somewhere between 34 and ($T_m=$) 53.2 °C, and in some cases it would not occur at all. The $T_n$ vary in different crystals of the same substance. .

- *Only crystal defects serve as the nucleation sites.* In other words, the nucleation is always heterogeneous as opposed to the homogeneous nucleation assumed to occur by a successful fluctuation in any point of ideal crystal lattice. The location of a nucleus can be foretold with a good probability when the crystal has a visible defect. Nucleation in sufficiently overheated / overcooled crystals can be initiated by introducing an "artificial defect" (a slight prick with a glass string). In such a case, a nucleus appears at the damaged spot. In cyclic L → H → L → H phase transitions the nucleus frequently appears several times at the same location.

- *The higher crystal perfection, the greater overheating or overcooling $\pm\Delta T_n = T_n - T_o$*. In brief, better crystals exhibit wider hysteresis, for $\pm\Delta T_n$ *is* hysteresis [2]. The correlation between the degree of crystal perfection and $\Delta T_n$ is confirmed in several ways. One, illustrated with the qualitative plot in Fig.4, is the $\Delta T_n$ dependence on the estimated quality of the single crystals. Different grades were assigned to sets of crystals depending on their quality, higher grades corresponding to higher crystal perfection. The grades were given on the basis of crystal appearance (perfection of the faces and bulk uniformity) and the way the crystals were grown (from solution, from vapor, rate of growth, temperature stability upon growing, etc.). For the solution-grown PDB crystals the graduation exhibited good correlation with the levels of $\Delta T_n$ within the grades 1 to 4. Two findings will be noted regarding $\Delta T_n$ in these crystals:
1. $\Delta T_n <1.9$ °C was not found even in the worst (grade 1) crystals, indicating the existence of a minimum (threshold) overheating required for nucleation to occur.
2. The most perfect among the solution-grown crystals were incapable of L→H changing, so it was L that

melted. There was no doubt in the availability of dislocation lines, individual vacancies, interstitial and foreign molecules
in those crystals, but they were not of the "proper" nucleation type. Yet, an "artificial defect", created by a slight prick at the temperature near $T_m$, immediately initiated the phase transition from the damaged spot.

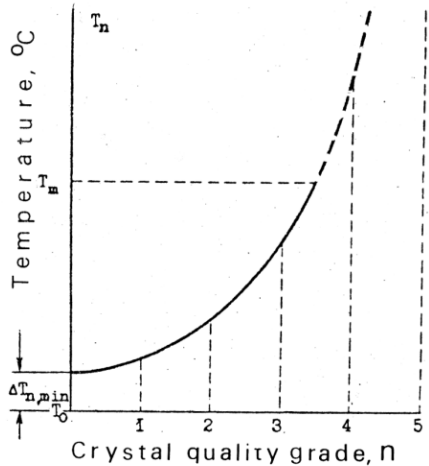

Fig. 4. The character of dependence of nucleation temperature $T_n$ on quality of single crystals; (L→ H). The quality is represented by number *n*: the higher estimated crystal perfection, the higher number *n*. No formation of a nucleus is possible until overheating exceeds some threshold value $\Delta T_{n,min}$. Overheating $\Delta T_n = T_n - T_o$ increases with *n*. Nucleation in "grade 4" crystals does not occur spontaneously, no matter how long they are stored just below melting point $T_m$. Yet, "grade 4" can be coerced into phase transition by a slight prick. No way has been found to induce a phase transition in "grade 5" crystals.

- *$T_n$ is pre-coded in the crystal defect acting as the nucleation site.* Fig. 5 shows the results of microscopic observation of PDB (L) single crystals upon slow heating. $T_n$ was recorded as soon as nucleation of the H phase was noticed. Growth of the H crystal was quickly stopped and the specimen was returned to the L phase. The cycle was repeated with the same crystal many times. Then the whole procedure was performed with another crystal. The experiments revealed that (a) a nucleus appears every time at exactly the same location, (b) the $T_n$ repeats itself as well, and (c) a particular $T_n$ is associated only with the particular nucleation site. A different $T_n$ is found in another crystal, also associated with the defect acting as the nucleation L site. Thus, *every crystal defect acting as a dormant nucleation site contains its own $T_n$ encoded in its structure*. If a crystal has more than one dormant nucleation site and is subjected to very slow heating, only one site with the lowest $T_n$ will be activated. Upon faster heating, the second nucleus of the second lowest $T_n$ may have time to be activated before growth from the first nucleus spreads over its location…and so on.

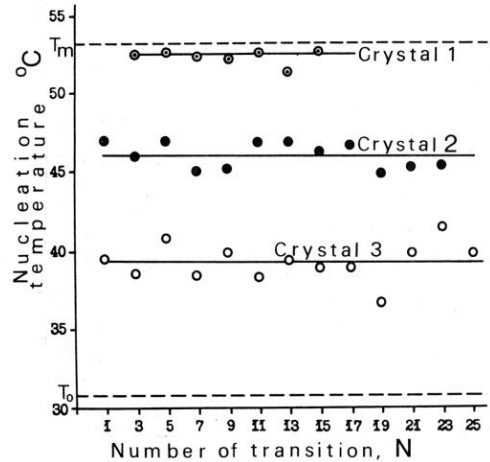

Fig. 5. Reiterative formation of a nucleus in small PDB single crystals. Reiteration of nucleation temperature $T_n$ when the nucleus repeatedly forms at the same site in the cyclic process. The procedure was as follows. Temperature of a microscope's hot stage with PDB single crystal was slowly raised. As soon as a nucleus of H phase became visible, $T_n$ was measured and the phase transition was immediately reversed. In the subsequent cycles the nucleus formed at the same site (as a rule, at a visible defect) and at the same $T_n$. The measurements for three single crystals are shown. Every particular site was associated with its particular $T_n$. N is ordinal number of a phase transition.

- *Orientation of the resultant crystal is also pre-coded in the crystal defect acting as the nucleation site.* The crystals of new phase in PDB phase transitions grow in random orientations. As long as a nucleus forms at the same lattice defect, the 3-D orientation and even the shape of the growing crystal repeats itself. Another nucleation site would produce a different orientation inherent exclusively to it. Thus, information on the orientation of the new crystal is pre-coded and stored in the structure of the crystal defect serving as the nucleation site. This information, however, remains dormant unless this defect is activated (by heating or cooling to the $T_n$ encoded in the defect).

- *The $T_n$ in the same crystal can be changed*. For example, in case of L→H phase transitions, the $T_n$, initially encoded in a nucleation site, can be "erased" by creating an "artificial defect", as mentioned earlier. The $T_n$ also goes down with every transition in the cyclic process L→H→L→H... if no precaution is taken to change the temperature slowly. As the number of cycles increases, the crystal deteriorates. Each successive transition originates from a new defect with a lower encoded $T_n$, but not lower than some threshold $\Delta T_{n,min}$ still above $T_o$.

- *A nucleation site is stable, although only to some extent*. A crystal can store a nucleation site, together with all the encoded nucleation information, for long time. The site can also withstand some influences such as moderate annealing, internal strains, and passage of interface. Suppose, the nucleation site is located at some point A. At the encoded temperature $T_{n,A} > T_o$ the site turns into a nucleus. If the transition L→ H is completed and reversed, the interface in the H → L transition will pass through the point A. This may leave the nucleation site A unaffected in a few successive cycles, so it will remain as such in the L→ H runs. Ultimately it will be destroyed and the transitions L→ H will not originate at point A any more. The observation is in accord with the reappearance of the same X-ray Laue pattern for several initial times noted in a long cyclical processes.

- *$\Delta T_n$ in a solid→solid phase transition is smaller than in a liquid→ solid phase transition*. At least, it was so in our experiments reproduced several times with different crystals. First, $\Delta T_n = | T_n - T_o |$ was measured in a PDB single crystal upon L → H and H → L transitions and was found to be in the 7 to 10 $^o$C range in both cases. Then the crystal was melted on the microscopic hot stage and permitted to cool down to room temperature. No crystallization occurred, demonstrating that $\Delta T_n$ (liq→sol.) > 32 $^o$C.

- *Neither nucleation nor growth is possible in "too perfect" single crystals*. In other words, crystal defects of *two* kinds are a necessary constituents of a solid-solid phase transition. Direct evidence of that was discovered when some vapor-grown PDB single crystals were found completely incapable of changing from L to H phase. Some of these crystals (grade 5 in Fig. 4) were very thin rod-like ones, prepared by evaporation into a glass tube, others were found on the walls of the jar where PDB was stored. Not only did they not change into the H phase upon heating, but it was also impossible to induce this change by means of the "artificial nucleation". Even after some of the crystals were badly damaged by multiple pricks with a needle, they still "refused" to change into the H phase at the temperature as high as only 0.2 $^o$C below $T_m$. Although all conditions for *starting* nucleation were thus provided, the nuclei could not grow: some additional condition was absent. That condition is a *sufficient concentration* of another type defects (see below). Anyhow, infeasibility of a homogeneous nucleation manifested itself unambiguously.

**4. Formation of a nucleus: a predetermined act, rather than a successful random fluctuation**

The conventional theory of nucleation in solids [50] considered nucleation of new phase to be the consequence of random heterophase fluctuations that give rise to formation of clusters large enough to become stable. The only change made in the formulae over nucleation in liquids and gases is an extra term representing the idea that nucleation barrier in solids is higher due to arising strains. The theory in question originally considered nucleation to be homogeneous (assuming equal probability for nuclei to appear at any point in the crystal), but had later to acknowledge that heterogeneous, *i.e.* localized at crystal defects, nucleation prevails. Therefore the theory was modified to take this into account. Heterogeneous nucleation is believed to occur at dislocation lines, foreign molecules and vacancies being present in real crystals in great numbers. In other respects the statistical-fluctuation approach has been left intact.

The real nucleation in a crystal is quite the opposite. The fact that $\Delta T_n$ (solid→solid) < $\Delta T_n$ (liquid→solid) runs counter to the theoretical premise that activation energy of nucleation in solids is higher due to the internal strains. Formation of a nucleus is a rare and reproducible act bound to a predetermined location. Homogeneous nucleation is impossible. As for heterogeneous nucleation, the formation of nuclei at dislocation lines or vacancies has to be ruled out as inconsistent with evidence. There are plenty of these defects in every real crystal and they change their position under the action of even the slightest internal strains, to say nothing of the strains caused by moving interfaces. Dislocation lines, vacancies, and foreign molecules cannot account for the observed *rarity, stability, and reproducibility* of a nucleus. They are too primitive to contain encoded individual information about $T_n$ and orientation of the resultant crystal.

**5. The structure of a nucleation site**

Finding the particular structure of a crystal defect serving as a nucleation site thus turned out to be the key to nucleation in solids. The solution is presented below. It inevitably involves an element of speculation, but only to link all elements of comprehensive evidence into a self-consistent and clear picture. It is qualitative, but more reliable than a detailed mathematical description of an idea that has not been verified.

Let us sum up the properties of these defects. There are few such defects in a good single crystal; they reside at permanent locations; they do not form spontaneously over long-term storage at any level of overheating / overcooling; they are stable enough to withstand rather strong influences; they are not quite as stable as macroscopic defects; they possess a memory large enough to contain individual information both on $T_n$ and nucleus orientation; they are capable of activating the stored information repeatedly; their structure permits nucleation without development of prohibitive strains.

Microcavities (cluster of vacancies) of some optimum size (Fig. 6) are, perhaps, the only type of crystal defect that meets all these requirements. An optimum microcavity (OM) eliminates the problem of the great strains that, possibly, prevent nucleation in a defect-free crystal medium. The OM consists of many individual vacancies and is therefore relatively stable and bound to a permanent location. There can only be a few such large-sized defects in a good single crystal, or even none. Yet, the defect in question is far from a macroscopic size, so it can be affected in some way to lose its nucleation function. Variations of its size and shape account for its capability of storing individual information on nucleation. (One additional vacancy can be attached to OM in many different ways, thus adding many new variants of "encoded" information).

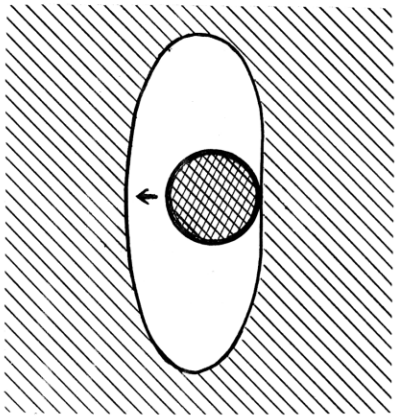

Fig. 6. Formation of a nucleus in an optimum microcavity. Not only there are no accompanied strains, but the activation energy of molecular relocation is especially low owing to attractive action of the opposite wall.

The relatively large size of OM has already been discussed. But OM also cannot be too large. Limited stability of OM is one argument in favor of such a conclusion. Now let us return to Fig. 6. It is true that a very large cavity eliminates the associated strains in the same way a smaller one does. In both cases a nucleus can grow freely on a cavity wall. The fact is, however, that a large cavity is equivalent to an external surface where no nucleation strains would be involved. But crystal faces do not facilitate nucleation, as it follows from the observation that nucleation in a "too perfect" crystal fails to occur both in the bulk and on the faces. Nucleation on the crystal faces obviously lacks some additional condition.

The idea on the nature of the lacking condition comes from the mode of molecular rearrangement at the contact interface (Sec. 2). There, molecular relocation from one side of interface to the opposite one proceeded under the attractive action of the latter, and this circumstance lowered the activation energy. Similarly, a strain-free nucleation with low activation energy is offered by a microcavity that is sufficiently narrow to facilitate the molecular relocation by an attractive action of the opposite wall. The gap must be of a molecular dimension. The particular width and configuration of this gap in a given OM may well be that same parameter which determines the encoded individual $T_n$. The detailed shape of OM may then be responsible for the encoded orientation of the nucleus. Thus, there is an intrinsic alliance between the molecular mechanisms of nucleation and growth. Both are based on the principle "relocation under attraction". When nucleus formation is completed, crystal growth takes over so naturally that these two stages of phase transition merge into a single unified process.

**6. Epitaxial nucleation**

The structure of a nucleation site does not require OR. Not rarely, however, OR is exhibited in certain phase transitions (setting aside those where OR has been assumed without verification). That does not mean these transitions occurred by a kind of "deformation" of the original phase or "displacement" of its atoms / molecules, as still frequently believed. They occur by nucleation and growth as well.

There are two circumstances of strict OR in phase transitions. One is layered crystal structures [3]. A layered structure consists of strongly bounded, energetically advantageous two-dimensional units − molecular layers − usually appearing in both phases. There the interlayer interaction is weak on definition. Since the layer stacking contributes relatively little to the total lattice energy, the difference in the total free energies of the two structural variants is small. This is why layered crystals are prone to polymorphism. Change from one polymorph to the other is mainly reduced to the mode of layer stacking. The layers themselves are only slightly modified by different layer stacking.

In practice, layered structures always have numerous defects of imprecise layer stacking. Most of these defects are minute wedge-like interlayer cracks located at the crystal faces as viewed from the side of layer edges. In such a microcavity there always is a point where the gap has the optimum width for nucleation. There the molecular relocation from one wall to the other occurs with no steric hindrance and, at the same time, with the aid of attraction from the opposite wall. In view of the close structural similarity of the layers in the two polymorphs, *this nucleation is epitaxial*. Orienting effect of the substrate (the opposite wall) preserves the orientation of molecular layers.

Another case of the epitaxial nucleation is when the unit cell parameters of the polymorphs are extremely close even in non-layered crystals, as in the Fe ferromagnetic phase transition [5]. This is also the reason for the rigorous OR in magnetization of polydomain structures where the "polymorphs" have identical crystal structure.

The kinetics of epitaxial phase transitions differs significantly from the non-epitaxial. Hysteresis $\Delta T_n$ in epitaxial phase transitions is very small. Thus, due to the abundance of wedge-like microcracks in layered crystals, there is no shortage in the nucleation sites of optimum size. At that, the presence of a substrate of almost identical surface structure acts like a crystallization "seed". Therefore only small overheating or overcooling is required in order to initiate and quickly complete this kind of solid-state reaction. Without a scrupulous verification, the phase transitions in question may be taken for (kinetics-free) "displacive", "instantaneous", "cooperative", "soft-mode", "second-order", etc.

( *Note*: Epitaxial nucleation is the cause of existence of polydomain structures due to formation of a nucleus in two or more equivalent positions if allowed by the substrate symmetry).

## 7. Interface motion: additional considerations.

Sec. 2 outlined how the interface moves forward by a "molecule-by-molecule" relocation at the molecular steps. The availability of some extra space at the steps to provide sufficient steric freedom for the relocation was noted as an important condition. This extra space eventually comes out on the surface and a new space must appear at the interface for the process to continue. Consequently, a phase transition can take place only in a *real* crystal with a sufficient concentration of vacancies and/or their clusters. The interface kinetics described in Section 7 offers strong support to that basic concept. Moreover, it made possible to gain insight into the intimate details of interface motion.

We deal with two types of nucleation: 3-D to initiate a phase transition (OM) and 2-D to initiate a new transverse layer to advance the interface in the direction of its normal. The normal velocity of the interface motion $V_n$ is controlled by the latter. The 2-D nuclei form only heterogeneously, like the OM do. There is a significant difference in their function. While only one OM is needed to start a phase transition, a sufficient concentration of appropriate defects is required to keep the interface moving. These defects are also microcavities, but smaller than OM, although not just individual vacancies. They will be called *vacancy aggregates* (VAs). One VA acts as a site for the 2-D nucleus only once and then moves to the crystal surface. The VAs may differ by the number and combination of constituent vacancies, as well as how close to the interface they are.

In the process of its motion the interface intersects the positions of VAs, which is equivalent to a flow of VAs onto the interface. Intensity of the flow depends on the concentration of VAs and is affected by VAs migration. Not all VAs of the flow can be effective, but only those with the activation energy of nucleation that is lower of a particular level. That level is determined by the overheating / overcooling $\Delta T_{2-D} = |T_{tr} - T_o|$, where $T_{tr}$ is the actual temperature of phase transition. One transition in a crystal "consumes" only a part of the available VAs, leaving repetition of the process possible.

Another major phenomenon of interface kinetics results from the fact that the interface motion is crystallization, and, consequently, a process of purification. The cause of the purification is obvious: attachment of a proper particle to the growing crystal is more probable than of a foreign one. (Due to the "repulsion" of foreign particles by a growing crystal, crystallization from liquid phase is utilized in practice for purification of substances). In this process any crystal defects are "foreign particles" as well. In particular, vacancies and VAs form a "cloud" in front of the moving interface, as sketched in Fig. 7. This phenomenon gives rise to the following effects:
  (a) intensification of the VAs flow to the interface, resulting in a faster interface motion;
  (b) intensification of coagulation of vacancies into VAs and VAs into larger VAs in the "cloud";
  (c) dissipation of the "cloud" with time due to migration of the defects towards their lower concentration, that is, away from the interface.

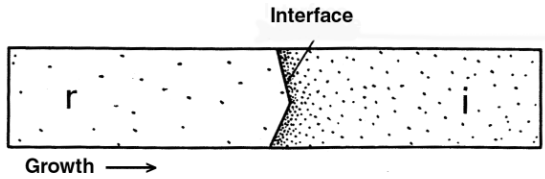

Fig. 7. Accumulation of crystal imperfections in front of moving interface. The phenomenon results from the fact that adding a proper particle to a growing crystal face is always more preferable than an improper. (Crystal defects such as vacancies or linear dislocations can also be considered "improper particles"). The well-known zone refining technique is based on the same principle.

The described mechanism of interface motion is responsible only for the basic phenomena of interface kinetics. These phenomena clearly exhibit themselves under certain idealized conditions: change in the specific volume upon the phase transition of the chosen object is small, the specimens are good small single crystals, the interfaces are flat, and the velocities of the interface motion are low. If these conditions are not met, the "basic kinetics" can be completely obscured by secondary phenomena considered in Sec. 9.

## 8. Experimental facts of interface kinetics

### (*1*) *No phase transition in defect-free crystal*

This experimental fact, already described in Sec. 3, is fundamental. If crystals are "too perfect" they do not change their phase state at any temperature. Even when OMs to start phase transition are purposely created, the interface motion cannot proceed, lacking a sufficient concentration of VAs. Obviously, the crystals in which the phenomenon was observed were still far from being ideal, still containing defects such as vacancies and linear dislocations. The defects that these crystals were lacking were VAs. A new VA is required to form 2-D nucleus on the interface every time the previous one exits on the crystal surface. If the VAs are present, but their concentration is lower than required for uninterrupted 2-D nucleation, there will be no phase transition. One would be wise to take this fact into account prior to undertaking a theoretical work on phase transitions in ideal crystal medium.

### (*2*) *Temperature dependence*

There is a strong dependence of the velocity V of an interface motion on temperature. There is a problem, however, with V(T) measurements due to the V dependence on the availability of VAs. Therefore, the experimental curve in Fig.8 should be considered as exhibiting only general qualitative features of temperature dependence. Every experimental point in the curve is the result of double averaging, first on all the transitions in each crystal, and then on different crystals. The curve shows that

- V=0 at $T_o$; ($\Delta T_{tr}=T_{tr}-T_o=0$). While $T_o$ is usually called "temperature of phase transition", it is the only temperature at which the phase transition is unconditionally impossible,
- V(T) is tangent to T-axis at $T_o$,
- V increases from zero as $T_{tr}$ moves away from $T_o$ up or down,
- left-hand part of the V(T) curve exhibits a maximum similar to that found in melt crystallization.

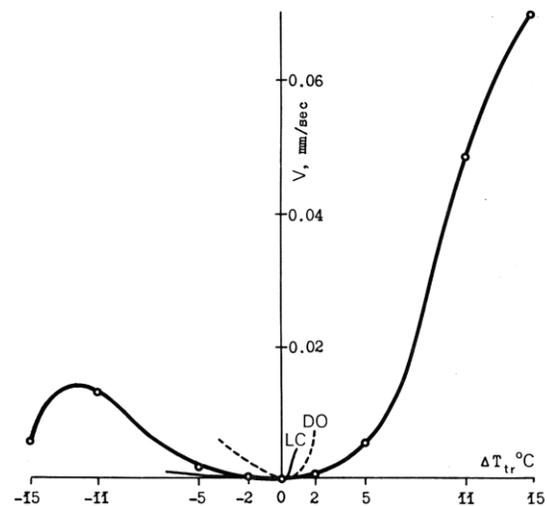

Fig.8. Velocity V of interface motion in PDB against $T_{tr}$. Each experimental point was the result of averaging. The curves marked with letters were drawn from literature data: LC [*Compt. Rend.* **248**, 3157 (1959)], and DO [*Docl. Acad. Nauk SSSR* **73**, 1169 (1960)] to demonstrate poor reproducibility typical for kinetics measurements.

Phenomenologically, two factors shape the curve in Fig. 8: driving force and absolute temperature. The former is determined by the difference between the free energies of the phases; V is zero at $T_o$ and increases as $T_{tr}$ is moved away from $T_o$ in any direction. The absolute temperature factor, on the other hand, affects V in one direction: the higher $T_{tr}$, the higher V. To the right from $T_o$ the two factors act in the same direction, causing progressive V increase; to the left from $T_o$ they act in opposite directions giving rise to the maximum.

### *(3) Hysteresis of interface motion*

From the two types of temperature hysteresis, the $\Delta T_{2-D}$ required to keep the interfaces moving is, as a rule, much smaller. In order to observe it, interface motion should first be stopped by setting $T_{tr} = T_o$, and then set in motion again by deviation from $T_o$. However slow and careful the last procedure is performed, one will find that some finite overheating / overcooling $\Delta T_{tr} \neq 0$ is required in order to resume interface motion. The situation in the vicinity of $T_o$ is shown schematically in Fig. 9. Interface can move only if $\Delta T_{tr}$ is greater than some threshold value. A phase transition is intrinsically a non-equilibrium phenomenon. Therefore, when one encounters the title "Non-equilibrium phase transitions", the question "Do equilibrium phase transitions exist?" would be justified.

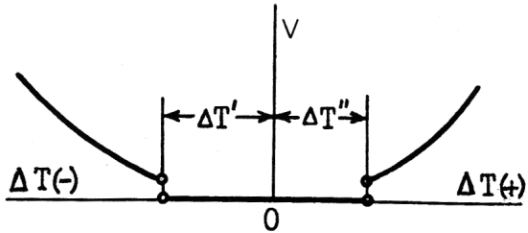

Fig. 9. Hysteresis of interface motion. The interface motion requires 2-D nucleation, and the latter requires overcoming some energy barriers. The sketch is to illustrate the experimental fact that $\Delta T$ lower of a certain minimum ($\Delta T'$ for cooling and $\Delta T''$ for heating) will not set an interface in motion. The phenomenon is similar to the hysteresis of 3-D nucleation.

(*4*) *Depletion of the reserve of lattice defects*

Interface motion in the small rod-like (0.22 x 5 mm) PDB single crystal shown in Fig. 10 was manipulated on a hot microscopic stage over a prolong time. The crystal, grown from a vapor phase, was of a rather high quality. By temperature control the interface (seen in the photograph) was moved back and forth many times without letting it to reach the crystal ends. After 10 to 15 cycles, the interface was moving slower in every successive cycle under the action of the same $\Delta T_{tr}$. Additional 10 to 15 cycles stabilized the interface at a fixed position: it became completely insensitive to temperature changes. In this state the specimen, consisting of the two phases divided by the interface, could be stored for days at room temperature 20°C, that is, about 11° lower than $T_o = 30.8°C$. After several days the resumed experiments revealed that the dependence $V = f(\Delta T)$ had been partially restored, but the same $\Delta T_{tr}$ produced much lower V. The initial V had not been regained even after several weeks.

The schematic in Fig. 10 illustrates the cause of the phenomenon. Cyclical movements of the interface over the length $\ell$ have a "cleaning" effect. While "consuming" some part of the VAs for the 2-D nucleation, the interface pushed other VAs out of the $\ell$ area. The interface completely stopped when the concentration of VAs, $C_{VA}$, fell below the critical level required for the renewable 2-D nucleation. A partial restoration of the motion capability after the long "rest" was due to VAs migration from the end regions to the "working" region $\ell$. This experiment makes the intrinsic irreproducibility of V quite evident. The velocity in the same specimen can differ by orders of magnitude under identical temperature conditions.

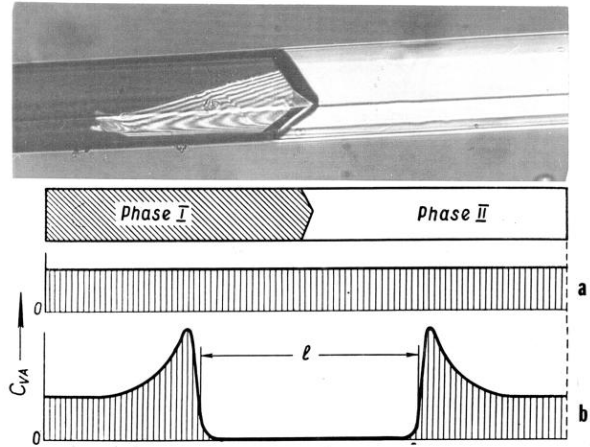

Fig. 10. Depletion of the reserve of lattice defects (VAs) needed for 2-D nucleation to keep the interface moving. The photograph shows the thin rod-like PDB single crystal with which the experiments were carried out. The drawings show what happens to the concentration of defects $C_{VA}$, being initially uniform (plot 'a'), after the interface was moved back and forth many times over the length $\ell$ (plot 'b').

(*5*) *Velocity V as a function of the number of transitions*

This experiment was similar to the described in the previous section, but with two differences: the crystals were not so perfect and V was measured in every interface run. The specimens were oblong PDB single crystals grown from a solution. Due to significant V scatter the measurements were averaged over 20 crystals. A region of 1 mm long was selected in the middle of a crystal and the time required for the interface to travel this distance was measured. The outside regions played a certain auxiliary role. Two microscopes with hot stages set at $T_1 = T_o + \Delta T$ and $T_2 = T_o - \Delta T$ were used in the measurements. The V dependence on the ordinal number N of transitions in the cyclic process was measured for a fixed $|\Delta T|$. The $V = f(N)$ plots are shown in Fig. 11. Only qualitative significance should be assigned to them.

A new finding is the maxima, and more specifically, their ascending side - because their descending side has been explained earlier. In general terms, a moving interface initially creates more VAs than it consumes, but the tendency is reversed after a number of successive transitions. In more detail, it occurs as follows. A moving interface accumulates a "cloud" of vacancies and VAs in front of it. If density of the vacancies in the "cloud" is sufficiently high, their merging into VAs creates more VAs than is expended on the 2-D nucleation. Considering that the number of vacancies in the region is limited, the consumption eventually prevails and V begins dropping. The recurrence of the whole effect after a long "rest" is due to migration of the vacancies from the end parts of the crystal.

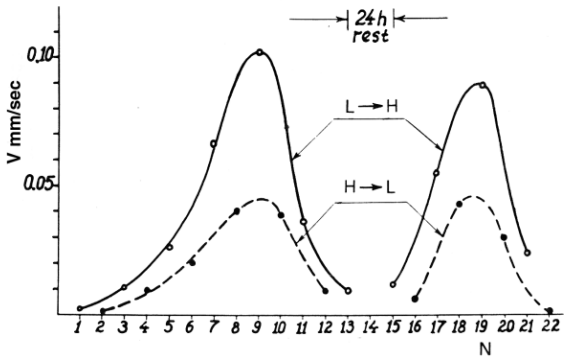

Fig. 11. Velocity of interface motion V in PDB single crystals as a function of number N of phase transitions in the cyclic succession L→H→L→H... Two phenomena are revealed: maxima of V and ability to show them again after a sufficiently long "rest".

*(6) Lingering in resting position*

If an interface moving with a speed $V_1$ at a fixed $\Delta T_{tr}$ was stopped by setting $\Delta T=0$, it tends to linger in the resting position once the initial $\Delta T_{tr}$ is restored. After some lag the interface leaves the resting position, but under a lower speed $V_2 < V_1$. The longer the resting time, the lower the $V_2$. Once resumed, its movement accelerates to approximately the previous steady $V_1$ level.

The diagram in Fig. 12 explains this peculiar behavior. Translational movement of interface pushes a "cloud" of crystal defects in front of itself. Its speed $V_1$ under isothermal conditions is controlled by the density of VAs in the cloud. The density, in turn, is controlled by the balance between accumulation and consumption of VAs. Holding the interface in one position allows the cloud to dissipate to the extent depending on the resting duration. In order to start moving again, the interface must now "dig" for VAs from the uniform distribution, leaving behind a "hole". As a result, $V_2 < V_1$, but approaches $V_1$ as the new "cloud" is accumulated.

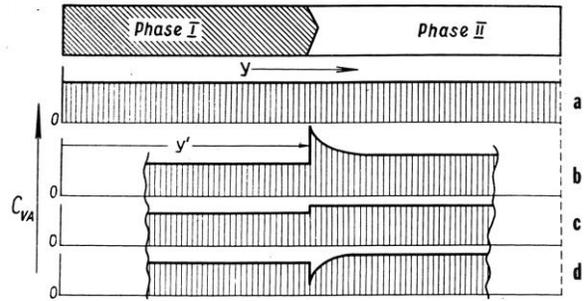

Fig. 12. The effects of moving and resting interface on the VAs concentration, $C_{VA}$.
  (a) The initial uniform distribution.
  (b) In the course of translational motion of the interface which is in the position y' at the moment.
  (c) After long rest in the position y'.
  (d) After the motion was resumed.

*(7) Memory of the previous position*

If after the procedure just described the reverse run immediately follows, the moving interface "stumbles" (is retarded spontaneously) exactly at the position where it was previously resting. The phenomenon is almost a visual proof that the "hole" shown in Fig. 12d really exists. It provides a compelling support to the concept of interface kinetics based on the flow of VAs on the interface.

*(8) Slower start upon repetition*

Using temperature control, it is possible to set up a cyclic process in which a single nucleus of H-phase will appear, grow to a certain small size, and then dissipate back to the L-phase. In such a process, growth of the H crystal in every subsequent cycle requires a longer time. The cause: the growing crystal consumes the available surrounding VAs for its 2-D nucleation, while the traveling distance is too short to accumulate a "cloud" of the defects. The concentration of VAs in the area is reduced with every successive cycle, giving rise to a lover V.

*(9) Acceleration from start*

Just after its nucleation, an H crystal grows very slowly. It takes some traveling distance for the interface to accelerate and attain a steady V level. A "cloud" of VAs, initially absent, is then accumulated. Eventually a kind of equilibrium between their accumulation and consumption is reached, producing (in a uniform crystal medium) a translational interface motion.

*(10) Acceleration induced by approaching interface*

When there are several H crystals growing from independent nucleation sites in the same L crystal, it can be easily seen that the rates of their growth vary in a wide range. Considering that the initial crystalline matter and $\Delta T_{tr}$ are equal for all the growing H crystals, this fact in itself is instructive in regard to kinetics of solid-state phase transitions: which of these rates does any existing theory account for? There is another phenomenon observed repeatedly: these rates are not quite independent of one another. In one instance, pictured in Fig. 13a, the crystal $r_1$ was almost not growing when a fast-growing interface from $r_2$ began approaching from the opposite end. The latter crystal noticeably activated the growth of the former when the two were still separated by as much as 1.5 mm. As the $r_2$ was coming closer, growth of $r_1$ sharply accelerated (Fig. 13b).

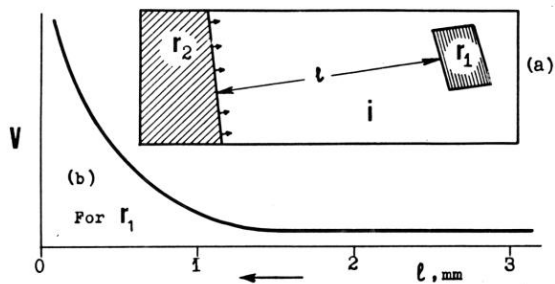

Fig. 13. Actuation and acceleration of growth caused by an approaching interface.
(a) A sketch picturing a real case when one crystal ($r_1$) was initially not growing under some overheating $\Delta T$ = const, and then was actuated by the approaching interface from $r_2$. The initial crystal is marked 'i'. (b) Change in the observed velocity V for $r_1$ (a qualitative representation, but the distances $\ell$ between $r_1$, and $r_2$ are close to real).

The crystal $r_1$ was initially not growing due to the lack of VAs in its vicinity. Transport of VAs from $r_2$ to $r_1$ has spurred its growth. The growth progressively sped up as the flow of VAs increased from the "cloud" driven by the approaching $r_2$. A plausible additional cause is the strains spreading from $r_2$ (faster-moving interfaces produce stronger strains). The strains set in motion the static VAs dwelling at some distance from $r_1$ and thus start to foster its growth even before it is approached by the "cloud" from $r_2$.

## 9. Revision of the activation energy concept

In experimental studies of kinetics of solid-state phase transitions the phase ratio was measured *vs.* time with the objective to find the "activation energy of phase transition $E_a$". The process of a phase transition was heterogeneous, but $E_a$ was interpreted as the energy barrier to be overcome in the process of a cooperative homogeneous rearrangement of one ideal crystal structure into another. The inconsistence of this approach is conspicuous. Considering that phase transitions between crystal states occur by 3-D nucleation and subsequent growth, there must be at least two activation energies: one for nucleation, the other for rearrangement at the interfaces. The latter process, in turn, involves two major stages: 2-D nucleation of molecular layers and molecular relocation at interfaces. The three basic activation energies that control the above three major stages of a solid-state phase transition are:

$E_a'$. *Activation energy of a 3-D nucleus formation*. The $E_a'$ depends on the particular structure (size- and configuration) of the lattice defect (OM) acting as the nucleation site. The nucleation temperatures encoded in these sites are different, therefore $E_a'$ is not a unique characteristic of a particular phase transition. Rather, it can be of any magnitude greater than $E'_{a,min}$ corresponding to the $\Delta T_{tr,min}$. Absence of even a single OM in the crystal is equivalent to $E_a' = \infty$. This leads to the conclusion that attempts to find the $E_a'$ characteristic of a given phase transition would be physically unsound. This activation energy has nothing to do with interface kinetics. Phase transition in a fine-crystalline powder exemplifies the case when the bulk rate of transition under changing temperature is governed exclusively by different $E_a'$ encoded in the individual particles.

$E_a''$. *Activation energy of 2-D nuclei formation on a molecular-flat interface*. $E_a''$ is not a fixed value either. It varies owing to structural differences (size and shape) of VAs acting as the nucleation sites. The VAs must be present in quantities and located near the interface in order that the latter be able to propagate. If this condition is not met, the phase transition (interface motion) will not be possible, which is equivalent to $E_a'' = \infty$. At moderate concentrations of VAs the interface motion is controlled more by the availability of VAs than the E" magnitudes. A lower speed of an interface motion at the same temperature is an example of interface kinetics governed by VAs availability. In the case of high VAs concentrations, when only a small part of the available VAs is "consumed" during interface motion, molecular relocation across the interface starts limiting the interface speed.

$E_a'''$. *Activation energy of molecular relocation at kinks of a contact interface.* As shown in Sec. 2, the process in question is a "stimulated sublimation". This activation energy is much lower then the previous two and rarely controls the linear kinetics.

## 10. Relationships between the controlling parameters

The complications and instabilities of interface kinetics are rooted in feedbacks. An interface needs certain conditions for its motion, but its motion affects these conditions. Flowchart below summarizes relationships between the parameters responsible for the interface kinetics controlled by VAs flow. After the foregoing discussion, the flowchart is self-explanatory even if it may seem cumbersome. Connections between the parameters should be traced from the bottom up following solid-line arrows. The feedbacks that turn the process into autocatalytic are shown by broken lines. The temperature effects are of two kinds. One is $\Delta T_{tr}$, which provides energy gradient for phase transition. The other is absolute temperature - the cause of molecular vibrations and other mobilities. The flowchart illustrates that (1) phase transition in an ideal crystal is not possible and (2) the phenomena of kinetics are complex, multiparameter and irreproducible in spite of the simplicity of the *contact* mechanism. Yet, the flowchart represents only the simplest case of slow interface motion in a good quality real single crystal and when the accompanying strains are sufficiently small not to create the additional complications described in the next section.

There is one more simplification in this flowchart, and it is significant: it does not reflect the phase transition latent heat which can dramatically affect its kinetics - up to explosion in some cases.

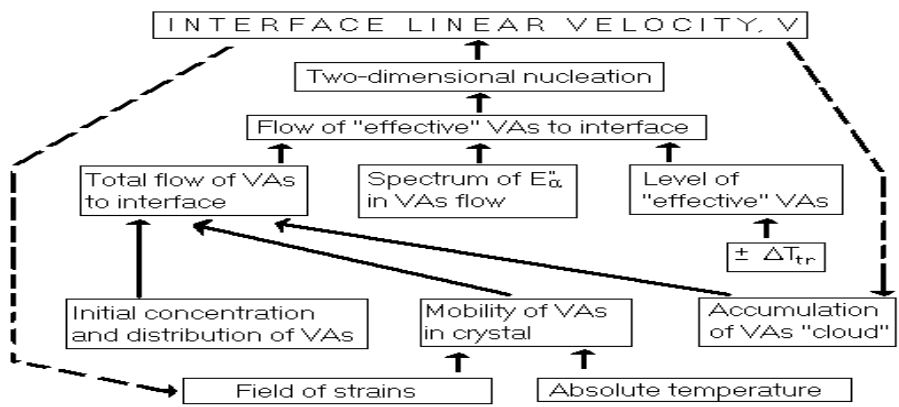

## 9. The "truly out of control" kinetics

If the previously described interface kinetics may seem "out of control", it still represents the simplest and most orderly case. A smooth advancement of a flat interface takes place only if certain precautions are taken: the specimen is a good small single crystal and $\Delta T_{tr}$ is low. It is also helpful if the specific volumes of the polymorphs are close and the crystal has a plate-like shape. Then the strains arising at the slowly moving interface can dissipate before damaging the original crystal medium. If these conditions are not favorable, the crystal growth loses visually orderly character. This disorderly morphology for a century delayed discovery of the underlying phenomenon presented in Sec. 2: growth of naturally-faced crystals. (Another cause of the delay was not using optical microscopy and transparent single crystals). The phase transition in most instances appears to the observer as a blurred thick "wave" rolling over the crystal and quickly completing the process, leaving behind a less transparent material. What kind of kinetics is that? The X-ray patterns reveal that a single crystal turns into a polycrystal. All facts taken together suggest that the interface *generates* multiple lattice defects, OMs, acting as the sites for 3-D nucleation immediately in front of itself. This is caused by the strains originating from the fast-moving interface. Because VAs are generated as well, the new growth proceeds quickly in both directions: toward the interface and out of it, creating new strains. Not having time for relaxation they again damage the adjacent lattice. This kinetics is based on the positive feedback:

INTERFACE MOTION → STRAINS →
GENERATION OF NEW NUCLEI JUST IN FRONT OF THE INTERFACE → GROWTH FROM THESE NUCLEI (INTERFACE MOTION) → STRAINS ...
and so on.

It should be noted that the term "interface" is used here only conditionally: actually, it is a rather thick heterophase layer. Such an interface can move with a very high speed. Fig. 14 shows an instance of a sudden change of orderly crystal growth to the kinetics based on the positive feedback. It exhibits itself as a local "explosion" with the higher speed of interface motion by one order of magnitude. Thus, in 'c' two different kinetics transparently manifest themselves in the same initial crystal.

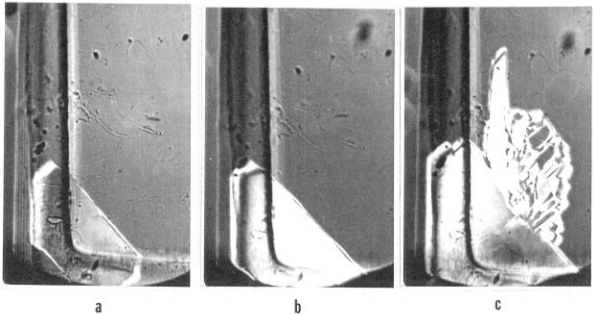

Fig. 14. Sharp conversion from a "quiet" interface kinetics based on consumption of available defects ('a' and 'b') to that based on generation of new defects by the strains spreading from the interface ('c'). Only one corner of the original crystal is shown. The conversion manifests itself as an "explosion" on the flat surface.

## 11. Solid-state recrystallization

The *contact mechanism* offers an insight into another solid-state reaction − recrystallization (migration of grain boundaries) of polycrystalline solids. It is hard to find any reason to assume one molecular mechanism for interface propagation in phase transitions and another for migration of grain boundaries. The grains in a polycrystal have the same crystal structure, but due to their random orientations the conditions at their boundaries are not different from those at phase transition interfaces. The grain boundaries do migrate. The difference is their driving forces. They are to minimize the grain surface energy and/or substitute a more perfect lattice for a less perfect one. They are much weaker, resulting in slower process. In all other respects it is the same crystal growth.

Recrystallization in itself is a large topic, a branch of physical metallurgy and some other applied sciences. The *contact* mechanism tells us in which direction a grain boundary moves, namely, from the grain where it has rational $(h,k,\ell)$ to the grain where it is irrational. It follows that a major component of the recrystallization driving force is the elimination of irrationally oriented boundaries. But there is more to it. The boundary will migrate only if the two neighboring grains have different orientations. The boundary between two grains of the same orientation will be either equally rational with the same $(h,k,\ell)$, or equally irrational. In such a case, no driving force to instigate the molecular relocation in one or the other direction exists: $E_a$ is same in either direction. The boundary remains still. In such a straightforward manner the *contact* mechanism accounts for one of the unexplained "recrystallization laws" which states that the boundary between two grains of the same orientation does not migrate [52].

One more example in the field of recrystallization. The following fact perplexed observers. Sometimes one part of the boundary between grains A and B migrates from A to B, while another part migrates from B to A (Fig. 15). In terms of the contact mechanism, the cause of the phenomenon is simply in the directions of the boundaries: in the former case, the boundary is a natural crystal plane in A and irrational in B, but the other way around in the latter case.

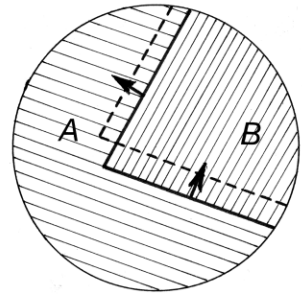

Fig. 15. In solid-state recrystallization, the A → B and B → A migration of the boundary between grains A and B can proceed simultaneously [53]. Dashed line indicates a subsequent position of the grain boundary.

## 12. Solid-state ferromagnetic transitions

Ferromagnetic phase transitions should be added to the list of solid-state reactions. Initially everyone believed that they are of the *second order* − a cooperative phenomenon with strictly fixed ("critical", "Curie") temperature of phase transition. In 1965 Belov wrote in his monograph "Magnetic Transitions" [54] that ferromagnetic and antiferromagnetic transitions are "concrete examples" of second-order phase transitions. But in 1970's the theorists were puzzled after a number of *first-order* ferromagnetic phase transitions were reported.. *It was not realized that a first-order phase transition meant nucleation and growth, and not a critical phenomenon* [1]. Since then the number of recognized first-order ferromagnetic phase transitions grew dramatically. They turned out to be of the first order even in the basic ferromagnetics − Fe, Ni and Co [7]. This process was accompanied by the increasing realization of structural changes involved. A new term *"magnetostructural"* transitions appeared and is being used to distinguish them from the "not structural".

There was no explanation why some ferromagnetic phase transitions are "accompanied" by structural change, and others do not. But explanation is simple, although controversial to many (not to this author): *all ferromagnetic phase transitions are "structural"*, meaning they always materialize by nucleation and crystal rearrangements at the interfaces, rather than cooperatively. Designations of phase transitions as second order are always superficial. Not a single sufficiently documented example, ferromagnetic or otherwise, exists. This is because a nucleation-growth phase transition represents the most energy-efficient mechanism, considering that it needs energy to relocate only one molecule at a time, and not the myriads of molecules at a time as a cooperative process requires.

Ferromagnetic phase transition is a crystal rearrangement *accompanied* by change of the magnetization. Moreover, no change in the state of magnetization is possible without the crystal reconstruction. This is a direct consequence of the natural principle that *the orientation of a spin is determined by the orientation of its atomic carrier*. Therefore, any reorientation of spins requires reorientation of their carriers. The only way to achieve that is replacing the crystal structure. This occurs by nucleation and interface propagation. Everything regarding the nucleation and growth is relevant and applicable to ferromagnetic phase transitions (in their epitaxial version). *All* ferromagnetic phase transitions are "magnetostructural". The term, however, is defective in the sense that it suggests existence of ferromagnetic phase transitions without structural change. Refer to [7] for details.

### 13. Magnetization by interface propagation.

Magnetization of polydomain crystals is a solid-state reaction as well, whether the driving force is temperature, pressure or applied magnetic field. The conventional theory does not explain why magnetization is realized by propagation of interfaces rather than cooperatively in the bulk. Once again: magnetization is not a spin reorientation in the same crystal structure; it requires turning the atomic / molecular spin carriers. The only way to turn the spin carriers is by crystal rearrangement. The mechanism of crystal rearrangements is nucleation and propagation of interfaces (in this case − polydomain twin boundaries).

Possibility of a cooperative magnetization "by rotation" is thus ruled out. Refer to [7].